\documentclass[conference]{IEEEtran}

\usepackage{amsmath,amsfonts}
\usepackage{graphicx}
\usepackage{textcomp}
\usepackage{xcolor}
\usepackage{enumitem}
\usepackage[linesnumbered,ruled,vlined]{algorithm2e}
\usepackage{pgfplots}
\pgfplotsset{compat=1.18}
\usepackage{subcaption}
\usepackage{booktabs}
\usepackage{threeparttable}
\usepackage{hyperref}
\usepackage{amssymb}
\usepackage{xparse}

\usepackage{multibib}
\newif\ifcameraready
\camerareadytrue   % show "new text" (blue) and hide old

\NewDocumentCommand{\camerareplace}{+m +m}{%
  \ifcameraready
    {\color{black}#2}%
  \else
    #1%
  \fi
}

\NewDocumentCommand{\cameraready}{+m +m}{\camerareplace{#1}{#2}}
\newcommand{\citeapp}{\cite}

\usepackage[textsize=tiny, textwidth=1.3cm, disable]{todonotes}

\usepackage{xspace}
\newcommand{\MoveOD}[0]{\textsc{MoveOD}\xspace}
\pgfplotsset{compat=newest,
        /pgfplots/ybar legend/.style={
        /pgfplots/legend image code/.code={%
        \draw[##1,/tikz/.cd,bar width=3pt,yshift=-0.2em,bar shift=0pt]
                plot coordinates {(0cm,0.8em)};},
},
}

\begin{document}

\title{MoveOD: Synthesizing Origin-Destination Commute Distribution from U.S. Census Data}

\author{
\IEEEauthorblockN{
Rishav Sen\IEEEauthorrefmark{1},
Jose Paolo Talusan\IEEEauthorrefmark{1},
Abhishek Dubey\IEEEauthorrefmark{1},\\
Ayan Mukhopadhyay\IEEEauthorrefmark{2},
Samitha Samaranayake\IEEEauthorrefmark{3},
Aron Laszka\IEEEauthorrefmark{4}
}

\IEEEauthorblockA{
\IEEEauthorrefmark{1}Vanderbilt University, Nashville, TN, USA\\
\IEEEauthorrefmark{2}William \& Mary, Williamsburg, VA, USA\\
\IEEEauthorrefmark{3}Cornell University, Ithaca, NY, USA\\
\IEEEauthorrefmark{4}Pennsylvania State University, State College, PA, USA\\[0.5em]
% \IEEEauthorrefmark{1}\{ammar.bin.zulqarnain,jose.paolo.talusan,ayan.mukhopadhyay,abhishek.dubey\}@vanderbilt.edu\\
% \IEEEauthorrefmark{2}kelly.napier@nashville.gov\\
% \IEEEauthorrefmark{3}\{corey.gens,jennifer.higgs,colleen.herndon\}@nashville.gov
}
}

\maketitle

\begin{abstract}

High-resolution origin–destination (OD) tables are essential for a wide spectrum of intelligent transportation systems applications, from modeling traffic and signal timing optimization to congestion pricing and vehicle routing.
However, outside a handful of data rich cities, such data is rarely available. We introduce \textsc{MoveOD}, an open-source pipeline that synthesizes public data into commuter OD flows with fine-grained spatial and temporal departure times for any county in the United States.
\textsc{MoveOD} combines five open data sources -- American Community Survey (ACS) departure time and travel time distributions, Longitudinal Employer–Household Dynamics (LODES) residence-to-workplace flows, county geometries, road network information from OpenStreetMap (OSM), and building footprints from OSM and Microsoft, into a single OD dataset.
We use a constrained sampling and integer-programming method to reconcile the 
OD dataset with data from ACS and LODES. Our approach involves: 
(1) matching commuter totals per origin zone, 
(2) aligning workplace destinations with employment distributions, and 
(3) and calibrating travel durations to ACS-reported commute times.
This ensures the OD data accurately reflects commuting patterns.
We demonstrate the framework on Hamilton County, Tennessee, where we generate roughly 150,000  synthetic trips in minutes, which we feed into a digital-twin simulator with a benchmark suite of classical and learning-based vehicle-routing {algorithms}. 
The {\textsc{MoveOD}} pipeline is an {end-to-end} automated system, enabling users {to easily apply it across the United States by giving only a county and a year; and it can be} adapted to other countries with comparable census {datasets}. {The source code} to a lightweight browser dashboard will be made publicly available.
\end{abstract}


\section{Introduction}
Granular origin-destination (OD) travel demand data is foundational for intelligent transportation networks and cyber-physical systems (CPS), enabling dynamic traffic management, predictive routing, and adaptive signal control systems~\cite{li2021learning, xu2021hierarchically, du2023safelight}. A central challenge in large-scale societal CPS is that feedback, control, and learning pipelines are increasingly data-driven, yet high-fidelity, representative real-world data is often unavailable~\cite{xu2024unified, wang2025real}. Without community-representative demand data, transportation CPS research frequently relies on datasets such as NYC taxi traces that do not reflect the target community, limiting the generalizability of resulting models and control policies.
However, collecting fine-grained, building-level OD data remains a persistent challenge. Traditional methods such as household travel surveys, Bluetooth MAC trackers, or mobile GPS traces are often sparse, noisy, or available only for specific travel modes. These sources struggle to generalize across urban, suburban, and rural environments, posing a significant challenge for the over 3,000 smaller transit agencies across the United States that lack integrated data pipelines~\cite{Perlman2022Emerging}.

Several publicly available nationwide datasets capture aspects of commuting, but none provides a complete distribution. The American Community Survey (ACS)~\cite{national2007using} and Longitudinal Employer-Household Dynamics (LODES)~\cite{abowd2004integrated} provide aggregate counts of residents and jobs by census unit, but lack departure times and fine-grained (i.e., building-level) locations. Building footprints from OpenStreetMap (OSM)~\cite{mooney2017review} or Microsoft (MSBF)~\cite{heris2020rasterized} pinpoint every structure but contain no occupancy or travel-demand information. Traffic datasets (e.g., INRIX) provide fine-grained demand aggregated over road segments, but cannot be decomposed into individual OD trips. Naively combining these marginal datasets (e.g., via iterative proportional fitting) may yield implausible results, such as hundreds of commuters assigned to a small residential building or every commuter departing in a single rush-hour spike.

\cameraready{}{These limitations motivate a synthetic approach. In this work, we introduce \textsc{MoveOD}, an open-source framework for synthesizing granular, time-dependent OD datasets by integrating heterogeneous data sources, including census marginals, employment flows, and road networks, with physical constraints such as building locations, spatial coherence, and time-dependent travel feasibility~\cite{kitchin2014data, thakuriah2017introduction}. While \textsc{MoveOD} produces synthetic data, it maintains statistical fidelity by matching observed marginals (residence-workplace locations, departure times, travel durations) while enforcing spatial coherence. The result is a CPS data and calibration layer that supports realistic closed-loop evaluation in simulation, optimization, and learning-based control, enabling downstream works to validate their pipelines against community-representative demand rather than geographically mismatched proxies.}
We focus specifically on daily commuting trips between residences and workplaces, allowing \textsc{MoveOD} to leverage spatial employment and housing density data while avoiding poorly measured recreational or freight travel.

\textbf{Our core contribution} is not a static dataset but a modular synthesis approach: \textsc{MoveOD} can generate OD datasets for any U.S. county and, in principle, for any region where similar marginal datasets are available. We release an interactive platform to generate commute OD data and demonstrate its utility on Hamilton County and Davidson County in Tennessee, US.

\section{Related Works}\label{sec:related}

\noindent\textbf{Open Benchmarks for Transportation AI}:
The past few years have witnessed remarkable growth in large-scale, openly licensed datasets that have catalyzed progress in traffic forecasting, routing, and autonomous driving research. Early sensor-centric research such as \textsc{METR‐LA} and \textsc{PEMS‐BAY}~\cite{li2017diffusion} sparked the first wave of spatio-temporal graph neural networks; subsequent releases expanded in size (e.g., \textsc{LargeST}~\cite{liu2023largest}) or semantic richness (e.g., \textsc{ScenarioNet}~\cite{li2023scenarionet}).
While indispensable, these datasets share two limitations that are critical for OD research: 
\emph{(i)} they focus on traffic states measured at fixed sensor locations, not on the population-level flows between origins and destinations;  
\emph{(ii)} coverage is typically confined to a few metropolitan areas, hindering work on domain transfer and equitable deployment across diverse cities.

\noindent\textbf{Origin–Destination Estimation}:
  
Several open projects attempt full synthetic generation.  
Open-PFLOW~\cite{kashiyama2017open} uses stochastic methods to allocate commuting trips to census microdata, providing disaggregate, time-resolved movement data that enables analysis of flows at different times of day.  However, its extensibility beyond Japan is limited, and the generation process can rely on private or commercial datasets for detailed spatial information, such as building locations and road networks.  The vehicle-network simulator of ~\cite{uppoor2013generation} produces microscopic trajectories for Vehicular Ad Hoc Network (VANET) studies, yet its demand model is tuned to a single European city and lacks transparency in demographic assumptions.

\noindent\textbf{Simulators for Transportation Systems}:
Comprehensive microsimulation platforms like SUMO~\cite{SUMO2018} and VISSIM~\cite{visim} are widely used in transportation research. However, they present significant limitations for deep reinforcement learning (DRL) applications in intelligent transit dispatching problems. These simulators are designed for detailed traffic analysis and model individual vehicle dynamics, signal timing, and microscopic interactions that create substantial computational overhead during DRL training, which typically requires thousands to millions of episodes. 

\noindent\textbf{Public-Transit and Multi-Modal Planning}:
Modern, data-driven transit planning uses OD matrices to size fleets and schedule services~\cite{cipriani2012transit, alonso2017demand}.  
Having finer-grained OD data improves the calibration of discrete-choice models ~\cite{guo2012discovering} and supports more accurate multi-modal assignment~\cite{guihaire2008transit}.  
However, most public transit simulations~\cite{liyanage2020agent, 10020973} assume a fixed commuter OD table obtained from proprietary surveys, limiting reproducibility and flexibility.

\section{Problem Statement and Data Sources}
\label{sec:problem}

% \noindent\textbf{Problem Statement}:
%
Our objective is to infer a fine-grained joint distribution of commute data at the level of individual buildings and minutes of the day. In our setting, origins are residences and destinations are workplaces.

\noindent\textbf{Spatial sets}
Let $\mathcal{G}$ be the set of census units (CUs)
and $\mathcal{B}$ be the set of all buildings. For each $g \in \mathcal{G}$, we define its building set
$$
  B_g = \{\,b \in \mathcal{B} : G(b)=g\},
$$
where 
$
  G: \mathcal{B} \to \mathcal{G}
$
maps a building $b$ to the census unit $G(b)$ that contains the building.

\noindent\textbf{Origin and destination notation} Each CU can serve as both an origin and a destination. Let $o \in O \subseteq \mathcal{G}$ denote origin CUs and $d \in D \subseteq \mathcal{G}$ denote destination CUs.
Specific buildings in the origin and destination CUs are written as
$
  b_o \in B_{o},  b_d \in B_{d},
$
where $b_o$ and $b_d$ are sampled from the set of buildings in their respective census units.

\noindent\textbf{Temporal distribution} Each day is indexed by minutes $t \in \mathcal{T}$.
The ACS departure-time blocks are
$
s \in \mathcal{S}
$
where $s = [t_l^s, t_u^s)$ represents an interval in hours (e.g., $[0,6)$ for ``Before 6 AM'' and $[6,7)$ for ``6--7 AM'').
To generate minute-level outputs, we sample a departure minute $m_o$ uniformly from the corresponding minute indices in the selected hour block, i.e.,
$m_o \in \{t_l^s,\ldots,t_u^s\}$.
Travel times are represented by bins $k \in \mathcal{K}$ (e.g., 0--5 minutes, 5--10 minutes, and so on).

\noindent\textbf{Commuters}
The total commuters per origin are represented as $N_o$ for $o \in O$.
The random vector describing a commuter’s trip is
$\bigl(B_o,\,B_d,\,M_o,\,M_d\bigr),$
where
$$
\begin{aligned}
  B_o &\in \mathcal{B} &&\text{is the origin building,} \\
  B_d &\in \mathcal{B} &&\text{is the destination building,} \\
  M_o   &\in \mathcal{T} &&\text{is the departure time from origin,} \\
  M_d   &\in \mathcal{T} &&\text{is the arrival time at destination.}
\end{aligned}
$$
A specific commuter is represented as $(b_o,b_d,m_o,m_d)$, and our goal is to estimate
$
  P\bigl(B_o=b_o,\,B_d=b_d,\,M_o=m_o,\,M_d=m_d\bigr),
$

\noindent\textbf{Notation conventions.}
We use lowercase symbols ($o,d,m_o,m_d$) as indices. The values $m_o$ and $m_d$ are minute-level realizations of the random variables $S$ (departure time) and $E$ (arrival time), respectively. We use uppercase symbols ($O,D,S,E$) for random variables, $n_{\cdot}$ for counts, and $p_{\cdot}$ for normalized probabilities.

\subsection{Public Data Sources}
Our framework relies exclusively on {publicly} available datasets to provide the spatial and temporal inputs needed for OD synthesis. 
Table~\ref{table:data_sources} provides a summary of the data sources used.
Below, we outline how we acquire and pre-process each dataset.

\noindent \textbf{Census unit (CU)}
To obtain the boundary of each CU, we download the U.S.\ Census Bureau’s TIGER/Line shapefiles for block groups (or another desired summary level) directly from the Census website. These open-source TIGER/Line files provide precise polygon boundaries for every block group in the United States.

\noindent\textbf{Commuter Origin-Destination}
The LEHD Origin--Destination Employment Statistics (LODES) provides a joint distribution of commuter counts for each origin--destination CU pair~\cite{lodes}. We represent
$
n_{o,d}
$
as the count of commuters with origin $o \in O$ and destination $d \in D$.

Then the estimated probability mass function for the $(o,d)$ pair, conditioned on the origin, is
$
p_{o,d}
=\frac{n_{o,d}}{\sum_{d'\in D} n_{o,d'}}.
$

\noindent \textbf{Departure Time Marginals (ACS Table B08302)}  
Let $\mathcal{C}_{\mathrm{dep}}$ be the ACS commuter records from Table B08302. We define
$n_{o,s}$ as the count of commuters originating at $o \in O$ with a departure time-block $s \in \mathcal{S}$.
By construction,
$
  \sum_{s \in \mathcal{S}} n_{o,s} \;=\; N_o,
$
i.e., the total weighted commuter count at origin $o$. We then set
$
  p_{o,s}
  \;=\; \frac{n_{o,s}}{N_o},
$
treating $p_{o,s}$ as the estimated origin–departure marginal, i.e., the population share leaving origin $o$ in each departure-time block $s$.

\noindent \textbf{Travel–Time Marginals (ACS Table B08303)}
Let $\mathcal{C}_{\mathrm{tt}}$ be the weighted set of travel-time records in ACS B08303. The travel-time bin $k \in \mathcal{K}$ indicates the time taken for a commuter to reach their destination. ACS provides the joint marginal only between origin and travel time. For each origin CU $o\in O$ and travel-time bin $k$, ACS B08303 reports a count
$n_{o,k}$. We obtain the empirical joint distribution
$
p_{o,k}
=\frac{n_{o,k}}{\sum_{k' \in \mathcal{K}}n_{o,k'}}.
$
which serves as our estimate of the travel-time distribution conditioned on origin.

\begin{table}[t]
\footnotesize
% \scriptsize
% \small
\centering
\caption{Data sources}
\vspace{-0.3cm}
\label{table:data_sources}
%\begin{tabular}{@{}llp{0.5\linewidth}@{}}
\begin{tabular}{lll}
\toprule
\textbf{Symbol} & \textbf{Type} & \textbf{Description} \\ 
\midrule
$\langle O, D \rangle$ & LODES & Origin-destination flow data \\
$\langle O, T \rangle$ & ACS B08302 & Departure time  by origin census unit \\
$\langle O, J \rangle$ & ACS B08303 & Travel time distribution by origin CU \\
$\mathcal{B}$ & OSM/MSBF & Building footprints (residence/workplace) \\
$\mathcal V$ & INRIX/OSM & Road speed data (hourly) \\
\bottomrule
\end{tabular}
% \vspace{-0.5cm}
\end{table}

\noindent \textbf{Tiered building selection.}
We start by identifying two subsets of $\mathcal{B}$.
$
  \mathcal{B}_{\mathrm{OSM}}
$
contains buildings with known locations and OpenStreetMap land-use tags $L(b)$ (residential, commercial, industrial, or other). OSM coverage can be incomplete.
The second subset is
$
  \mathcal{B}_{\mathrm{MSBF}},
$
from Microsoft's GlobalMLBuildingFootprints data, which provides broad building coverage but does not include land-use tags.
If a census unit has no building data in either source, we use the CU centroid $b_{\mathrm{centroid}}(g)$ as the fallback building location.

For each origin CU $o \in O$, the candidate building set is

$$
B_o = \begin{cases}
\{b \in \mathcal{B}_{\text{OSM}} : G(b)=o, L(b)=\text{res}\}, & \text{if} \neq \varnothing, \\
\{b \in \mathcal{B}_{\text{MSBF}} : G(b)=o\}, & \text{else if} \neq \varnothing, \\
\{b_{\text{centroid}}(o)\}, & \text{otherwise}
\end{cases}
$$

For each destination CU $d \in D$, the candidate destination-building set is

$$
B_d = \begin{cases}
\{b \in \mathcal{B}_{\text{OSM}} : G(b)=d, L(b)=\text{com}\}, & \text{if} \neq \varnothing, \\
\{b \in \mathcal{B}_{\text{MSBF}} : G(b)=d\}, & \text{else if} \neq \varnothing, \\
\{b_{\text{centroid}}(d)\}, & \text{otherwise}
\end{cases}
$$

\noindent \textbf{Road speeds.}
All road segments $z \in \mathcal Z$ have a speed $v \in \mathcal V$. The speed is determined using:
\begin{itemize}[leftmargin=*]
  \item \textbf{OpenStreetMap (OSM) defaults:} {Provides the road network and each road segment’s} speed limit, or a standard average speed based on the {type of road}.
  \item \textbf{INRIX data (proprietary, optional):} Provides hourly historical road speeds using average measurements from roadside sensors.
\end{itemize}

{We use INRIX speed values for roads where INRIX data is available; for other roads, we use speed values from OSM.}
This hybrid approach yields a granular travel-time profile $\tau_z(t)$ for road segment $z \in \mathcal{Z}$ at all times $t \in \mathcal{T}$. The availability of real-time road speeds makes synthetic data generation and calibration more accurate.

\section{Approach}
\label{sec:methods}

Having outlined the dataset requirements, we next detail the statistical integration process used to generate the commute OD representation.

\noindent \textbf{Bayesian Decomposition.}
Our target is to generate the complete joint distribution
$P(O,D,S,E)$, where $O$ is the origin CU, $D$ is the destination CU, $S$ is departure time, and $E=S+K$ is arrival time. By the chain rule,
$
  P(O,D,S,E)
  = P(E\mid S,D,O)\,P(S\mid D,O)\,P(D\mid O)\,P(O).
$
We simplify with the following assumption:
Departure-time blocks $S$ depend only on the origin:
    $P(S\mid D,O)=P(S\mid O)$.
Thus
$
  P(O,D,S,E)
  = P(E\mid S,D,O)\,P(S\mid O)\,P(D\mid O)\,P(O).
$
We estimate each of these marginals as follows:
\begin{itemize}[leftmargin=*]
  \item $P(O)$ from the CUs in a county.
  \item $P(D\mid O)$ from LODES, and then assigning each commuter an origin and destination building.
  \item $P(S\mid O)$ from departure-time marginals (ACS B08302).
  \item $P(E\mid S,D,O)$ by combining road-network travel times with ACS travel-time marginals (Table B08303).
\end{itemize}

Combining these steps yields a fully specified $P(O,D,S,E)$ that is consistent with available public distributions and can be sampled to produce building- and minute-level synthetic OD trips.
\cameraready{}{Keeping $P(D\mid O)$ and $P(S\mid O)$ intact anchors the synthetic data to observed spatial and temporal commuting patterns.}
These data sources give us four anchors:

\begin{enumerate}[leftmargin=*]
\item Origin-to-destination counts $n_{o,d}$.
\item 
Building location sets $B_{o}$ and $B_{d}$ represent the residential and commercial buildings, respectively, in every CU.
\item Conditional departure-time distribution $p_{o,s}$.
\item Conditional travel-time distribution $p_{o,k}$.
\end{enumerate}

The methodology below turns these anchors into minute-level, building-specific OD assignments $\mathcal A=\{(b_o,b_d,m_o,m_d)\}$, which are then calibrated against the travel-time distribution in ACS Table~B08303.

\noindent \textbf{1. Spatial Sampling}\label{sec:spatial}
Within each census unit, we assign origin and destination buildings uniformly to each of the $n_{o,d}$ commute trips, prioritizing tagged buildings over untagged ones. 
For each origin CU, we select an origin building $b_o\in B_o$; for destination CU $d$, we select $b_d\in B_d$.

\noindent \textbf{2. Departure-Time Assignment}\label{sec:departure}
Given an origin CU $o$ and departure-time block $s\in\mathcal{S}$ (with minute bounds $t_l^s, t_u^s$), we sample a minute-level departure time uniformly within the corresponding minute range:
$
  m_o \sim \mathrm{Unif}\bigl(\{t_l^s,\,\ldots,\,t_u^s\}\bigr).
$
This preserves the origin and departure-time marginals by construction.

\noindent \textbf{3. Arrival at Destination}\label{sec:raw_time}
For each sampled tuple $(b_o,b_d,m_o)$, we compute:
\begin{itemize}
  \item route length $\ell_{o,d}=\mathrm{graph\mbox{-}distance}(b_o,b_d)$,
  \item shortest-path travel time $\tau_{o,d}(m_o)=\mathrm{duration}(b_o,b_d\mid m_o)$,
  \item implied speed $v_{o,d}(m_o)=\ell_{o,d}/\tau_{o,d}(m_o)$.
\end{itemize}
Define
\[
  \Omega=\{\tau_{o,d}(m_o)\}
\]
over sampled trips. The arrival minute is
\[
  m_d = m_o + \tau_{o,d}(m_o).
\]
Hence the initial granular OD dataset is
\[
  \mathcal{A}=\{(b_o,b_d,m_o,m_d)\}.
\]

\noindent \textbf{Initial OD Dataset}\label{sec:initial_assign}
The dataset $\mathcal A$ satisfies ACS origin totals, LODES origin-destination marginals, and ACS departure-time marginals by construction, but still requires calibration to ACS travel-time marginals.

\noindent \textbf{Mean Road Speed Shift}\label{sec:mean_speed_shift}
To correct for any bias in travel-time reporting in ACS Table B08303, we perform a mean road-speed shift. Self-reported travel times usually overestimate actual travel times because of rounding errors and the inclusion of parking or waiting times~\cite{stopher2007household, peer2014over}.

Let $N=\sum_{o\in O}N_o$ be the total number of commuters. The mean travel time from the initial OD dataset is
\[
  \bar\tau^{\mathrm{init}} = \frac{1}{N}\sum_{\tau\in\Omega}\tau.
\]

Let $\bar\tau^{\mathrm{ACS}}$ be the ACS mean commute time from Table B08303, computed from travel-time bin midpoints. We use the multiplicative speed-shift factor
\[
  \psi = \frac{\bar\tau^{\mathrm{init}}}{\bar\tau^{\mathrm{ACS}}},
\]
and adjust road-segment speeds by $v_r' = \psi\,v_r$.

\section{Calibration}\label{sec:calibration}

The goal is to calibrate $\mathcal{A}$ to match ACS travel-time marginals while preserving origin-destination and origin-departure marginals.

\noindent \textbf{Inputs for a fixed origin CU $o\in O$}
\begin{enumerate}
  \item $m_{o,d,s}\in\mathbb{Z}_{\ge 0}$: commuters in $\mathcal A$ from origin $o$ to destination $d$ in departure block $s\in\mathcal S$.
  \item $N_o\in\mathbb{Z}_{>0}$: total commuters whose home CU is $o$.
  \item $p_{o,d}\ge 0$, $\sum_{d\in D}p_{o,d}=1$: destination-CU distribution.
  \item $p_{o,s}\ge 0$, $\sum_{s\in\mathcal S}p_{o,s}=1$: departure-time distribution (ACS B08302).
  \item $p_{o,k}\ge 0$, $\sum_{k\in\mathcal K}p_{o,k}=1$: travel-time-bin distribution (ACS B08303).
  \item $\pi_{o,d,s,k}\in[0,1]$: fraction of commuters in cell $(o,d,s)$ whose travel time falls in bin $k$.
\end{enumerate}

\noindent \textbf{Decision variables}
\begin{enumerate}
  \item $a_{o,d,s}\in\mathbb{Z}_{\ge 0}$: calibrated commuters assigned to $(d,s)$.
  \item $\eta_k^+,\eta_k^-\ge 0$: slacks for travel-time bin $k\in\mathcal K$.
  \item $\zeta_{o,d,s}^+,\zeta_{o,d,s}^-\ge 0$: slacks from initial counts $m_{o,d,s}$.
\end{enumerate}

\subsection{Integer linear program}
\noindent\textbf{Objective:}
\[
\min\; \alpha\sum_{k\in\mathcal K}(\eta_k^+ + \eta_k^-)
+ \beta\sum_{(d,s)\in D\times\mathcal S}(\zeta_{o,d,s}^+ + \zeta_{o,d,s}^-)
\]

\noindent\textbf{Subject to:}
\begin{enumerate}
  \item $\displaystyle \sum_{(d,s)\in D\times\mathcal S} a_{o,d,s}=N_o$
  \item $\displaystyle \forall d\in D:\ \sum_{s\in\mathcal S}a_{o,d,s}=N_o\,p_{o,d}$
  \item $\displaystyle \forall s\in\mathcal S:\ \sum_{d\in D}a_{o,d,s}=N_o\,p_{o,s}$
  \item $\displaystyle \forall k\in\mathcal K:\ \sum_{(d,s)\in D\times\mathcal S}\pi_{o,d,s,k}a_{o,d,s}+\eta_k^- -\eta_k^+=N_o\,p_{o,k}$
  \item $\displaystyle \forall(d,s)\in D\times\mathcal S:\ a_{o,d,s}+\zeta_{o,d,s}^- -\zeta_{o,d,s}^+=m_{o,d,s}$
\end{enumerate}

Constraints (1)--(3) enforce destination and departure marginals exactly. Constraint (4) matches travel-time marginals up to slacks $\eta_k^{+},\eta_k^{-}$, and constraint (5) keeps calibrated counts close to the initial OD dataset.

The calibrated OD counts $\{a_{o,d,s}\}$ are mapped back to buildings by resampling within each $(o,d,s)$ cell and sampling departure minutes inside each block $s$, then setting arrival time as
\[
m_d = m_o + \tilde\tau_{o,d}(m_o),
\]
where $\tilde\tau_{o,d}(m_o)$ is the mean-speed-shifted travel time.

\section{Experimental Setup \& Results}\label{sec:exp_results}

We demonstrate that the pipeline behaves as intended by applying it to Davidson and Hamilton counties in Tennessee, U.S. With more than 336,000 commuters, Davidson County is the largest county in Tennessee. We later use the Hamilton County OD dataset in multiple benchmarks. All experiments were performed on a 32-core, 5.2 GHz, 32 GB RAM Unix machine.

For the $\alpha, \beta$ hyperparameters in the calibration step, we sweep both over $[0,1]$ in increments of $0.1$. Experimentally, $\alpha = 1$ and $\beta = 1$ provide the smallest combined gap between the initial OD dataset $\sum_{(d,s)} (\zeta^{+}_{o,d,s} + \zeta^{-}_{o,d,s})$ and the ACS travel-time distribution $\sum_{k\in \mathcal{K}} (\eta^{+}_k + \eta^{-}_k)$.

% \begin{figure}[h] 
% \centering \includegraphics[width=0.42\textwidth]{images/compare_origin_dest.png} \caption{\textsc{MoveOD} preserves the conditional destination distribution for each origin census unit, ensuring that the synthetic workplace assignments match the empirical residence to work flow proportions reported in LODES.
% } \label{fig:origin_dest_comp} \end{figure}
\begin{figure}[t]
  \centering
\begin{tikzpicture}
  \begin{axis}[
    width=8.8cm,
    height=4cm,
    ybar,
    bar width=3pt,
    enlarge x limits=0.05,
    ymin=0, ymax=15,
    ylabel={Trip Distribution (\%)},
    xlabel={Departure Time Bin},
    symbolic x coords={
      00:00--04:59,05:00--05:29,05:30--05:59,
      06:00--06:29,06:30--06:59,07:00--07:29,
      07:30--07:59,08:00--08:29,08:30--08:59,
      09:00--09:59,10:00--10:59,11:00--11:59,
      12:00--15:59,16:00--23:59
    },
xtick=data,
xticklabels={
  {12a\\5a},
  {5a\\5:30a},
  {5:30a\\6a},
  {6a\\6:30a},
  {6:30a\\7a},
  {7a\\7:30a},
  {7:30a\\8a},
  {8a\\8:30a},
  {8:30a\\9a},
  {9a\\10a},
  {10a\\11a},
  {11a\\12p},
  {12p\\4p},
  {4p\\12a}
},
x tick label style={font=\scriptsize, rotate=0, align=center},
    % xtick=data,
    % xticklabels={
    %     {12am to 5am},
    %     {5am to 5:30am},
    %     {5:30am to 6am},
    %     {6am to 6:30am},
    %     {6:30am to 7am},
    %     {7am to 7:30am},
    %     {7:30am to 8am},
    %     {8am to 8:30am},
    %     {8:30am to 9am},
    %     {9am to 10am},
    %     {10am to 11am},
    %     {11am to 12pm},
    %     {12pm to 4pm},
    %     {4pm to 12am}
    %   },
    % x tick label style={font=\footnotesize, rotate=45,anchor=east},
    yticklabel style={/pgf/number format/.cd,fixed,precision=0},
    legend style={
      at={(0.86,0.57)},
      anchor=south,
      legend columns=1,
      /tikz/every even column/.append style={column sep=1cm},
    },
    grid=major,
    major grid style={line width=0.3pt, draw=gray!50},
  ]

    % Generated (Overall)
    \addplot coordinates {
      (00:00--04:59,3.7) (05:00--05:29,2.9) (05:30--05:59,3.9)
      (06:00--06:29,8.3) (06:30--06:59,9.4) (07:00--07:29,14.1)
      (07:30--07:59,14.2) (08:00--08:29,10.9) (08:30--08:59,6.5)
      (09:00--09:59,6.5) (10:00--10:59,3.1) (11:00--11:59,1.3)
      (12:00--15:59,7.2) (16:00--23:59,7.9)
    };

    % Census (Overall)
    \addplot coordinates {
      (00:00--04:59,3.7) (05:00--05:29,2.9) (05:30--05:59,3.9)
      (06:00--06:29,8.3) (06:30--06:59,9.4) (07:00--07:29,14.1)
      (07:30--07:59,14.2) (08:00--08:29,10.9) (08:30--08:59,6.5)
      (09:00--09:59,6.5) (10:00--10:59,3.1) (11:00--11:59,1.3)
      (12:00--15:59,7.2) (16:00--23:59,7.9)
    };

    \legend{Generated, Census}
  \end{axis}
\end{tikzpicture}
\vspace{-0.4cm}
  \caption{\MoveOD calibrates the departure times for all origin census units to align with ACS departure times (B08302).}
  % \Description{MoveOD calibrates the departure times for all origin census units to align with ACS departure times (B08302).}
  \label{fig:dep_time_matching}
\end{figure}

\begin{figure}[t]
  \centering
\begin{tikzpicture}
  \begin{axis}[
    width=8.8cm,
    height=4cm,
    ybar,                                % turn on bars
    bar width=2.5pt,                      % slimmer bars
    enlarge x limits=0.05,              % give some extra space at edges
    ymin=0, ymax=22.5,
    ylabel={Trip distribution (\%)},
    xlabel={Travel Time bins (minutes)},
    symbolic x coords={$<$5,5-9,10-14,15-19,20-24,25-29,30-34,35-39,40-44,45-59,60-89,90+},
    xtick=data,
    xticklabels={
      {$<5$},
      {5--9},
      {10--14},
      {15--19},
      {20--24},
      {25--29},
      {30--34},
      {35--39},
      {40--44},
      {45--59},
      {60--89},
      {90+}
    },
    x tick label style={font=\scriptsize, rotate=0, align=center},
    % --- group style to lay out 3 bars side by side ---
    % group style={
    %   group size=3 by 1,    % three plots in one row
    %   horizontal sep=4pt,   % separation between bars
    % },
    % --- legend on top, horizontal, no title ---
    legend style={
      at={(0.86,0.4)},
      anchor=south,
      legend columns=1,
      /tikz/every even column/.append style={column sep=1cm},
    },
    grid=major,
    major grid style={line width=0.3pt, draw=gray!50},
  ]

    % Initial
    \addplot coordinates {
      ($<$5,  8.6) (5-9, 16.0) (10-14,18.9) (15-19,17.45)
      (20-24,14.55) (25-29,10.35) (30-34,6.85) (35-39,4.0)
      (40-44,2.0) (45-59,1.35) (60-89,0.05) (90+,0.0)
    };

    % Calibrated
    \addplot coordinates {
      ($<$5,  7.5) (5-9, 14.0) (10-14,16.7) (15-19,17.25)
      (20-24,15.0) (25-29,10.3) (30-34,8.75) (35-39,4.7)
      (40-44,3.0) (45-59,2.5) (60-89,0.25) (90+,0.02)
    };

    % ACS
    \addplot coordinates {
      ($<$5,  2.4) (5-9, 9.5)  (10-14,14.6) (15-19,21.0)
      (20-24,18.45) (25-29,8.5) (30-34,13.35) (35-39,2.8)
      (40-44,2.6) (45-59,4.1) (60-89,1.45) (90+,1.35)
    };

    \legend{Initial, Calibrated, ACS}
  \end{axis}
\end{tikzpicture}
\vspace{-0.4cm}
  \caption{Travel time distributions: initial commuter assignment (Initial), calibrated commuter assignment (Calibrated), and ACS travel time distribution (Table B08303).}
  % \Description{Travel time distributions: initial commuter assignment (Initial), calibrated commuter assignment (Calibrated), and ACS travel time distribution (Table B08303)}
\vspace{-0.4cm}
  \label{fig:travel_time_matching} 
\end{figure}

\noindent \textbf{Validating the marginals.}
% Figure \ref{fig:origin_dest_comp}
% overlays the ACS modified LODES origin-departure distribution. The Jaccard similarity scores, $J(A,B) = |A \cap B| / |A \cup B|$, measuring the overlap between destination sets for each origin, exhibits a mean value of $1.0$ for all origin CUs. This Jaccard similarity score indicates near-perfect structural preservation. 
Figure \ref{fig:dep_time_matching} overlays the ACS departure-time histogram with the synthetic departures; the two curves coincide, confirming that the block-level departure times are reproduced exactly.
Figure~\ref{fig:travel_time_matching} compares the distribution of commute travel times across the initial commuter assignment (in blue), the calibrated commuter assignment (in red), and the ACS Table B08303 distribution (in brown). The calibrated synthetic data are more closely aligned with the ACS distribution than the initial synthetic data, both in the left tail (up to 15 minutes) and in the right tail (20 to 90 minutes), while preserving the mode and overall shape. However, the calibrated dataset slightly underestimates the proportion of typical travel times (15--19 minutes) compared to ACS. The calibration process thus improves fidelity to observed census patterns, but some discrepancies remain around the mean of the distribution. Overall, the results indicate strong structural preservation and successful calibration for most commuters.

Principally, the Hamilton County experiment shows that \MoveOD (i) preserves the origin-destination and origin-departure marginals, (ii) improves alignment with ACS travel-time marginals through calibration, and (iii) exposes a tunable parameter to account for survey bias in travel times.

\noindent \textbf{Time Complexity.}
The main cost scales linearly with the number of trips $M$. Spatial sampling and departure-time assignment are $O(M)$; travel-time computation is $O(M\cdot \log Y)$, where $Y$ is the number of nodes. Speed adjustment is $O(M+Z)$, with $Z$ road edges. Overall runtime is $O(M\cdot \log Y+Z)$. \MoveOD runs in 22 minutes for $\sim$336K commuters (Davidson County, TN) and 14 minutes for $\sim$150K commuters (Hamilton County, TN).
 
\noindent \textbf{Space Complexity.}
Memory use is $O(N)$ for commuter assignments, $O(B)$ for building tables, $O(G\cdot (S+K))$ for ACS marginals, and $O(Z+Y)$ for the road network. Here, $N$ is the number of commuters, $B$ is the number of buildings, $G$ is the number of census units, $S$ is the number of departure-time bins, $K$ is the number of travel-time bins, $Z$ is the number of road segments, and $Y$ is the number of network nodes. Total space complexity is $O(N+B+G\cdot (S+K)+Z+Y)$.

\begin{figure*}[t]
  \centering
  \captionsetup[subfigure]{justification=centering,skip=1pt}
  \begin{subfigure}[b]{0.24\linewidth}
    \includegraphics[width=\linewidth]{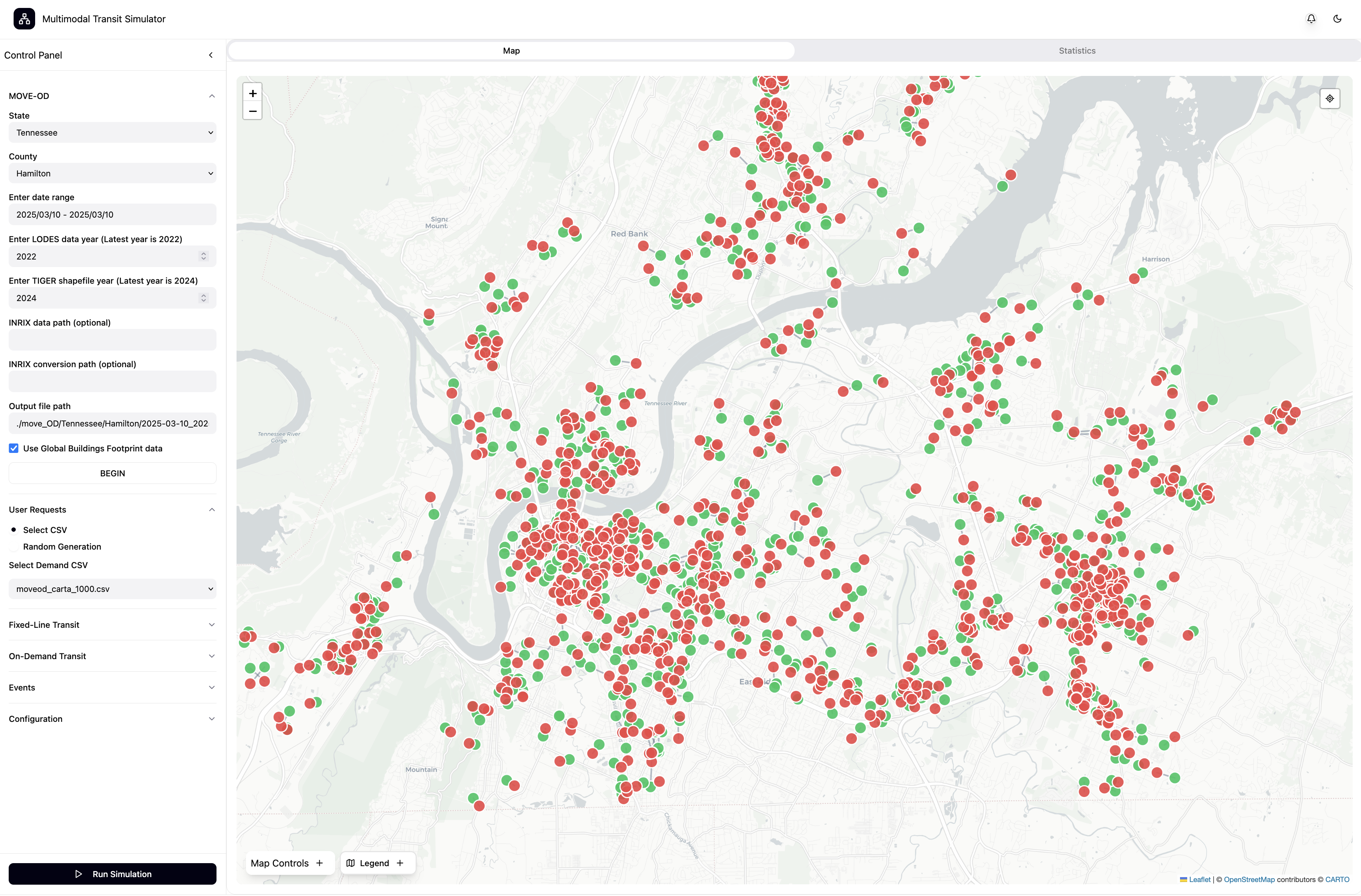}
    \caption{Synthetic MoveOD Data: Residence to Workplace}\label{fig:01_moveod}
  \end{subfigure}\hfill
  \begin{subfigure}[b]{0.24\linewidth}
    \includegraphics[width=\linewidth]{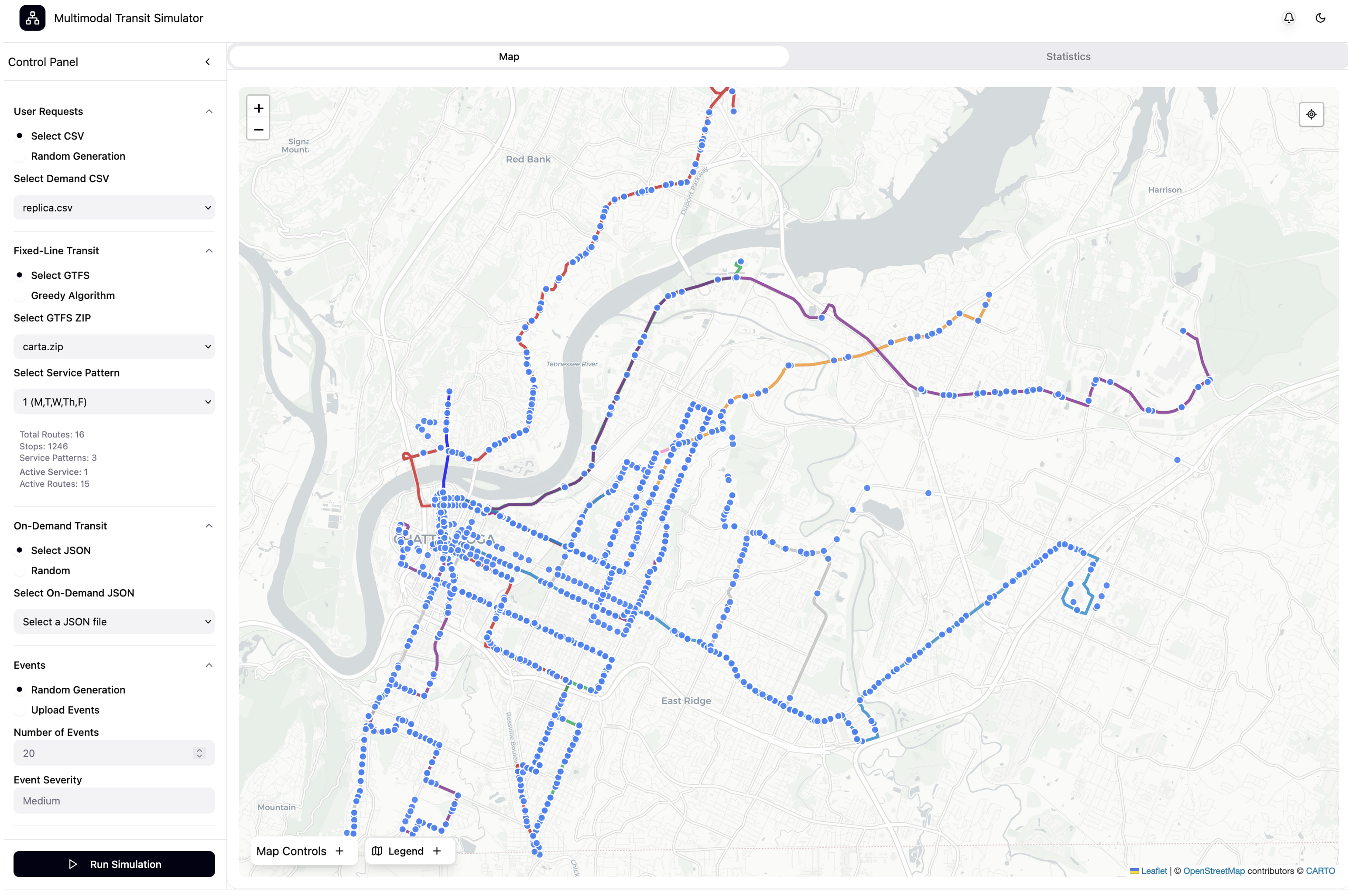}
    \caption{Public Transit routes and stops across 
    Hamilton County}\label{fig:02_gtfs}
  \end{subfigure}\hfill
  \begin{subfigure}[b]{0.24\linewidth}
    \includegraphics[width=\linewidth]{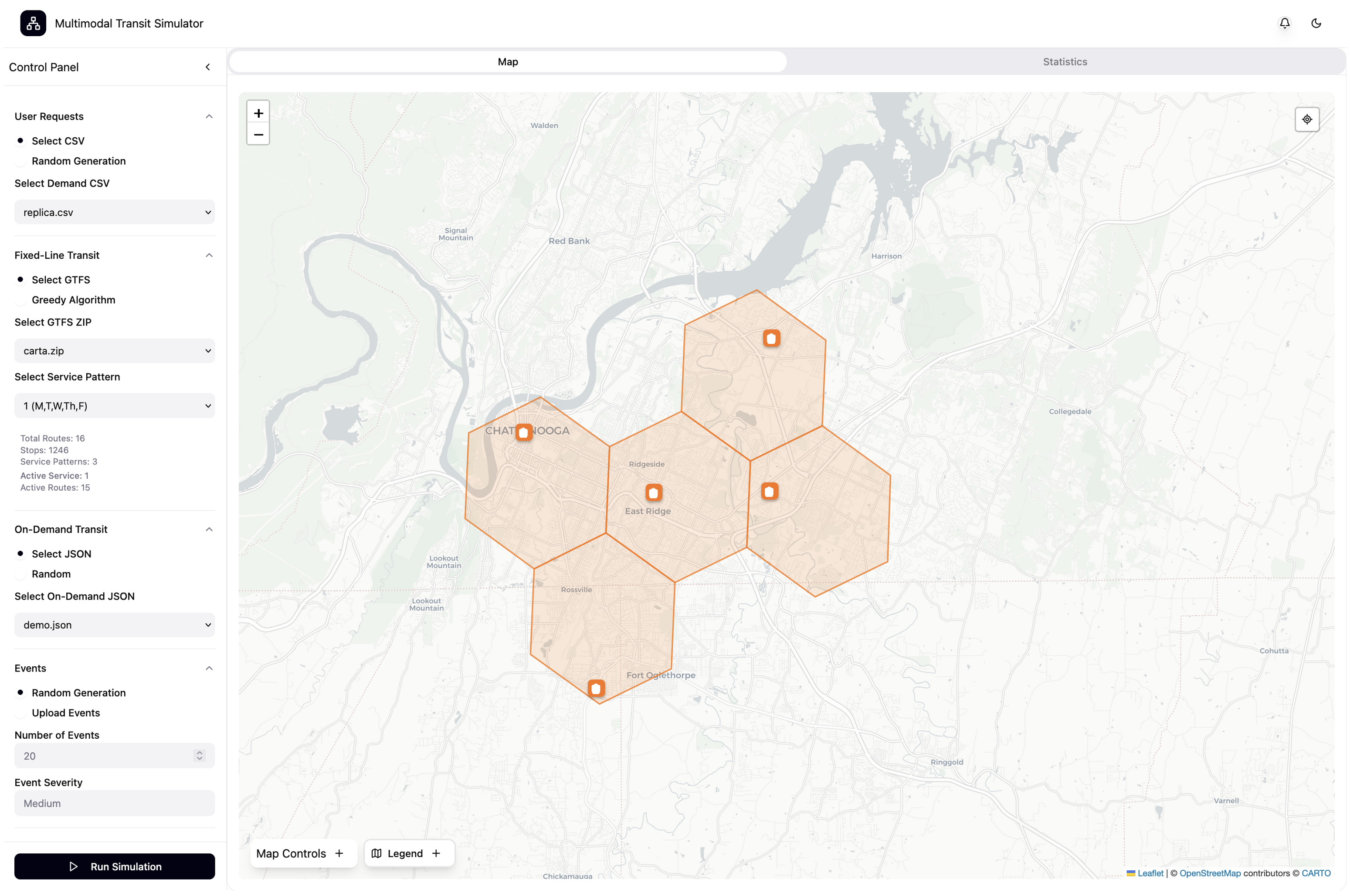}
    \caption{Sample On-Demand Service Zones and Depot}\label{fig:03_ondemand}
  \end{subfigure}
  \begin{subfigure}[b]{0.24\linewidth}
    \includegraphics[width=\linewidth]{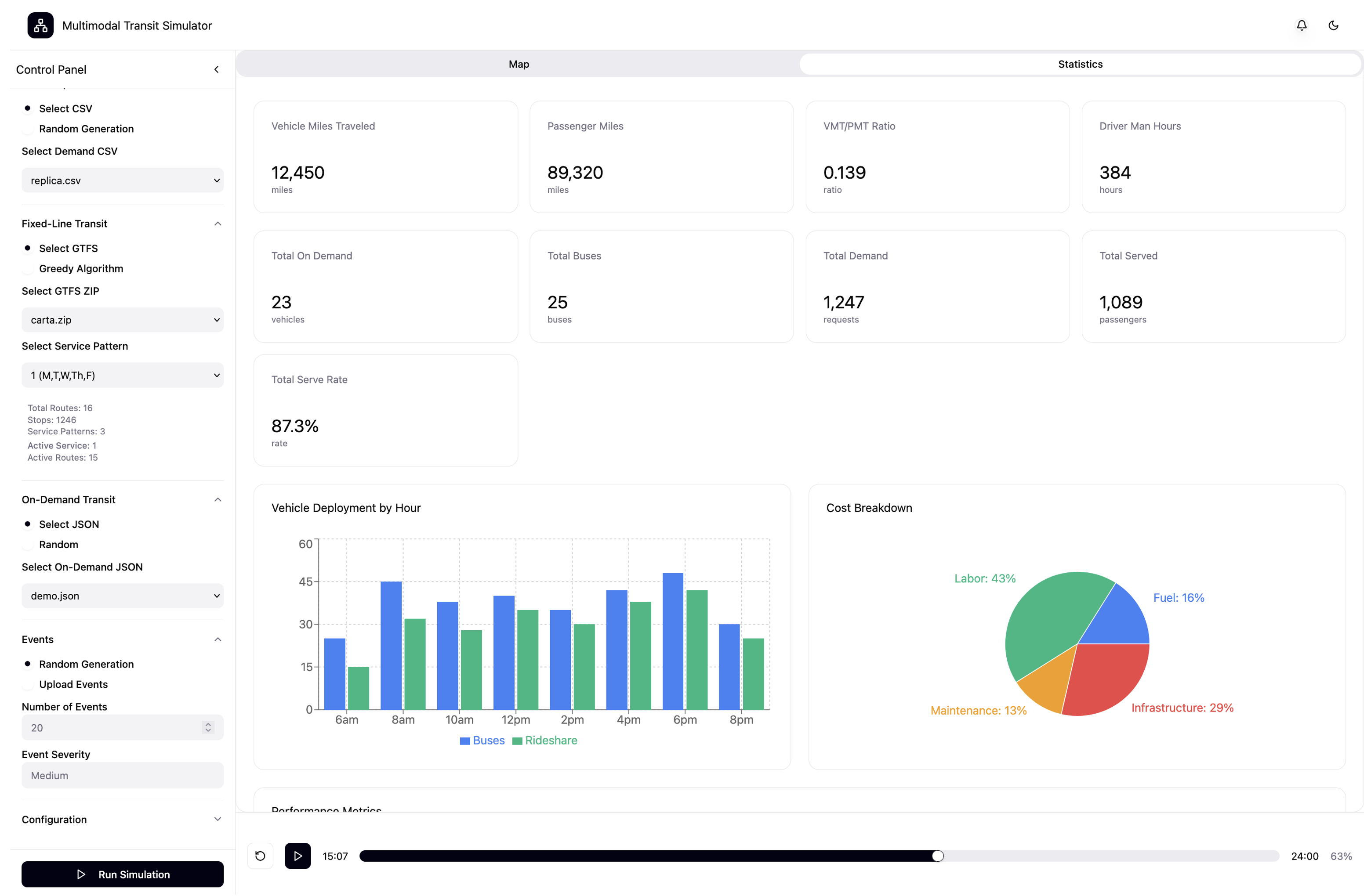}
    \caption{Metrics for multimodal transit given MoveOD requests}\label{fig:05_performance}
  \end{subfigure}
  \caption{\MoveOD data generation and digital twin dashboard. (a) Residential building (green) and workplace (red) locations. (b) CARTA public transit routes from public GTFS. (c) Sample On-demand configuration, where each hexagon represents service zones for respective on-demand services. (d) Performance metrics for the given configuration.}
  % \Description{MoveOD data generation and digital twin dashboard. (a) Residential building (green) and workplace (red) locations. (b) CARTA public transit routes from public GTFS. (c) Sample On-demand configuration, where each hexagon represents service zones for respective on-demand services. (d) Performance metrics for the given configuration.}
  \label{fig:digital_twin_figs}
\end{figure*}

\subsection{Interactive Interface and Digital Twin}

\noindent\textbf{Data Generation.}
\textsc{MoveOD} uses a minimalist web interface, Fig.~\ref{fig:01_moveod}, where a user selects the state, county, range of dates to synthesize, the LODES release year, and an optional INRIX road-speed feed. The number of points has been truncated for visibility.
Other inputs, such as ACS tables, OSM, MSBF, and CU boundary files, are retrieved automatically.
Pressing \texttt{BEGIN} returns ready-to-use calibrated OD tables, building metadata, and shapefiles.
% within minutes.

\noindent\textbf{Simulation.} The digital twin can ingest different user-demand requests, including those generated by \MoveOD, to simulate a range of transit configurations, from fully fixed-line service and on-demand service to multimodal systems that use optimization algorithms to identify the best mode for each user at any time. Figures~\ref{fig:02_gtfs} and~\ref{fig:03_ondemand} show how the twin can load existing General Transit Feed Specification (GTFS) and/or on-demand configurations. The tool allows operators or analysts to switch between systems and adjust configurations to evaluate performance for the user demand shown in Fig.~\ref{fig:05_performance}.

\noindent\textbf{Open Source.}
A browser-based, front-end tool will be made available at 
\cameraready{}{\url{https://github.com/scope-lab-vu/moveOD}.}
This repository lets users deploy a visual dashboard that allows them to choose a state, county, date, and optional spatial or temporal filters, then generate calibrated OD data in minutes.

\section{Discussion and Conclusions}\label{sec:disc}

\MoveOD delivers building- and minute-level commuter flows by fusing LODES OD distributions, ACS departure-time and travel-time distributions, and building footprints from OpenStreetMap and Microsoft Building Footprints. Unlike prior efforts that consider either where or when people commute, \MoveOD matches both spatial and temporal dimensions to ground-truth marginal distributions, while remaining fast enough to apply to any U.S. county.

\noindent\textbf{Typical applications.}
OD data show where people commute for work, which is one of the primary drivers of peak traffic conditions. OD data can be used for applications such as (i) predicting county-level traffic flows, (ii) public-transit design, and (iii) efficient road-network design.
Moreover, the data generation framework is designed to be flexible: it can ingest any macro-level movement dataset and enhance spatial resolution to building-level origins and destinations. 
If temporal distributions are available, the process further refines the OD dataset with time-aware commuter assignments. This adaptability allows integration with diverse data sources, such as crowd-sourced GPS trajectories or emerging Internet of Things (IoT) mobility streams.

\noindent\textbf{Limitations and outlook.}
\MoveOD currently generates OD data only for weekday commute trips.
Future work will add weekend and non-commute trips, account for weather and seasonal effects, and ingest real-time traffic speed data.

% \begin{acks}
\section*{Acknowledgment}
% Results in this paper were obtained with the Chameleon Testbed supported by the National Science Foundation.
This work is based upon the work sponsored in part by the National Science Foundation under grants 2238815 and 1952011. Any opinions, findings, and conclusions or recommendations expressed in this material are those of the author(s) and do not necessarily reflect the views of the NSF. 
Results reported in this paper were obtained using the Chameleon testbed, which is supported by NSF.
% \end{acks}

% \bibliographystyle{ACM-Reference-Format}
\bibliographystyle{IEEEtran}
\bibliography{main}

\newpage
\appendix

\section{Benchmark Experiments}
\label{sec:benchmarking}

In this section, we explain how we generate benchmark vehicle routing problem (VRP) instances from our calibrated OD data and evaluate different algorithms. The VRP~\citeapp{eksioglu2009vehicle, toth2014vehicle} is a classic optimization problem that asks how to route a fleet of vehicles as efficiently as possible to serve a set of locations. It is a motivating use case because it closely mirrors real-world challenges in public transit and ride-sharing.

\subsection{Benchmark Vehicle Routing Problems}
\label{sec:instance}

The calibrated commuter set
$\mathcal A=\{(b_o,b_d,m_o,m_d)\}$ contains hundreds of thousands of trips, which usually exceed
the computational limits of routing algorithms.
Therefore, we draw a random sample $\tilde{\mathcal A}$ of size $n$ without replacement:

Because every commuter in $\mathcal A$ is equally likely
to be chosen (probability $a/|\mathcal A|$),
the expected number of sampled trips that originate in census unit $o$ is

$$
  \mathbb E\!\bigl[\,|\tilde{\mathcal A}\cap\mathcal A_o|\,\bigr]
  \;=\;
  \frac{a}{|\mathcal A|}\,|\mathcal A_o|,
$$
which preserves (in expectation) the relative magnitudes of
origin-specific flows while making the instance tractable.
\footnote{If a deterministic size per origin is required,
one may draw $a_o=\lfloor a\,|\mathcal A_o|/|\mathcal A|\rfloor$ trips from each $\mathcal A_o$; we use the simpler
global random sample without replacement in our experiments.}

\noindent \textbf{VRP instance construction.}
Each sampled tuple $(b_o,b_d,m_o,m_d)\in\tilde{\mathcal A}$ is converted
into a customer of a Vehicle Routing Problem (VRP).

\begin{itemize}
\item \textbf{Depots:}  
      One or more operating centers per county; vehicles start and finish their routes at the depots. In our case, the depot is chosen as the centroid of all CUs in the dataset (for both training and test datasets).

\item \textbf{Customer nodes:}  
      Each commuter induces a pickup--delivery pair
      $(b_o,\,b_d)$ where  
      $b_o\in B_{o}$ is the $n$-th commuter’s
      home building and  
      $b_d\in B_{d}$ is that commuter’s
      workplace building. Both locations are taken directly from building footprints.

\item \textbf{Demand:} Each request is for a single passenger.

\item \textbf{Cost:}  
      Edge cost $c_{ij}$ is the great-circle distance between
      buildings $i$ and $j$ (or the corresponding driving distance/time if a road network is supplied).

\item \textbf{Time windows:}  
      The pickup node $b_o$ inherits the departure time $s$.

\end{itemize}
The objective is to serve all demand while minimizing both vehicle travel distance and passenger travel distance.
This sampling strategy yields routing test beds that are both
computationally manageable and statistically faithful to the {complete}, calibrated OD demand.

\begin{table*}[htbp]
\centering
\caption{CPDP Algorithm Performance (30 Vehicles{~with 5-Passenger Capacity})}
\label{tab:cpdp_results}
%\begin{tabular}{l|c|c|c|c|c|c|c}
\begin{tabular}{lccccccc}
\toprule
\textbf{Algorithm} &
\textbf{VMT$\;\!\downarrow$} &
\textbf{PMT$\;\!\downarrow$} &
\textbf{VMT/PMT$\;\!\downarrow$} &
\textbf{Empty$\;\!\downarrow$} &
\textbf{Coverage$\;\!\uparrow$} &
\textbf{Vehicle Utilization$\;\!\uparrow$} &
\textbf{Routes$\;\!\uparrow$} \\
\midrule
LKH-3                     & \textbf{2,255.16} & 1,455.74 & \textbf{1.55} & \textbf{40.9} & \textbf{100} & 100 & 30 \\
Clarke–Wright Savings     & 2,622.84          & 1,455.74 & 1.80          & 49.9          & \textbf{100} & 100 & 30 \\
Large Neighborhood Search & 2,960.06          & 1,455.74 & 2.03          & 56.2          & \textbf{100} & \hphantom{1}\textbf{67} & \textbf{20} \\
Genetic Algorithm         & 3,032.40          & 1,455.74 & 2.08          & 52.0          & \textbf{100} & \hphantom{1}\textbf{67} & \textbf{20} \\
Simulated Annealing       & 3,239.05          & 1,455.74 & 2.23          & 55.1          & \textbf{100} & 100 & 30 \\
Insertion Heuristic       & 3,281.13          & 1,455.74 & 2.25          & 55.6          & \textbf{100} & \hphantom{1}\textbf{67} & \textbf{20} \\
Ant Colony Optimization   & 3,281.13          & 1,455.74 & 2.25          & 55.6          & \textbf{100} & \hphantom{1}\textbf{67} & \textbf{20} \\
POMO                      & 3,296.79          & 1,455.74 & 2.26          & 55.8          & \textbf{100} & \hphantom{1}\textbf{67} & \textbf{20} \\
Attention Model           & 3,302.41          & 1,455.74 & 2.27          & 55.9          & \textbf{100} & \hphantom{1}83 & 25 \\
\bottomrule
\end{tabular}
\\[1ex]
\begin{minipage}{0.95\textwidth}
\footnotesize
Arrows indicate performance direction: $\downarrow$ {means }lower is better, $\uparrow$ {means }higher is better.  
VMT – Vehicle Miles Traveled; PMT – Passenger Miles Traveled; Empty – percentage of distance traveled without passengers; Coverage – percentage of passengers served; Vehicle Utilization – percentage of available vehicles used; Routes – number of vehicle routes{}.
\end{minipage}
\end{table*}

\subsection{Benchmark VRP Algorithms}
\label{sec:algs}

We apply a diverse selection of benchmark algorithms to the generated VRP instances, including classical heuristics, meta-heuristics, and deep reinforcement learning (DRL) approaches.
All implementations are the authors’ reference code, executed with default hyper-parameters unless noted otherwise.

\subsubsection*{Classical constructive heuristics}

\begin{enumerate}
\item{\textbf{Insertion}} \citeapp{randall2022insertion} The canonical cheapest insertion algorithm {constructs vehicle routes} by repeatedly inserting a customer whose marginal cost is minimal.
\item\textbf{Clarke–Wright} \citeapp{pichpibul2013heuristic} 
 A vehicle routing heuristic that starts with individual routes for each customer and iteratively merges pairs of routes. At each step, it selects the merge that maximizes the travel-cost savings.
 \end{enumerate}

\subsubsection*{Meta-heuristics}
\begin{enumerate}
\item\textbf{Simulated Annealing} \citeapp{czech2002parallel} A single–tour search that
      accepts uphill moves with a probability governed by a decaying temperature schedule.
\item\textbf{Genetic Algorithm} \citeapp{baker2003genetic} Population-based evolutionary
      search with order-crossover and inversion mutation.
\item\textbf{Ant Colony Optimization} \citeapp{rizzoli2007ant}  A constructive meta-heuristic in which artificial “ants’’ deposit pheromone on promising edges, reinforcing good routes over time.
\item\textbf{Large Neighborhood Search} \citeapp{dumez2021large} Repeatedly destroys and repairs parts of a solution; here we use the classical Shaw destroy and regret-2 repair operators.
\end{enumerate}

\subsubsection*{Exact / heuristic hybrid}
\begin{enumerate}
\item\textbf{LKH-3}\citeapp{helsgaun2017extension} Helsgaun’s state-of-the-art Lin–Kernighan implementation which combines variable-opt moves, 1-tree lower bounds and potent kick-starts.
\end{enumerate}

\subsubsection*{Deep-RL policies}

\begin{enumerate}
\item\textbf{Attention:}
Kool et al.~\citeapp{kool2018attention} propose an attention-based model trained with REINFORCE for learning heuristics to solve combinatorial optimization problems, specifically routing problems like the Traveling Salesman Problem and Vehicle Routing Problem. % (VRP).
\item\textbf{POMO:}  Policy Optimization with Multiple Optima algorithm \citeapp{kwon2020pomo} is a deep reinforcement learning {approach} designed to address instability in policy gradient methods, particularly when applied to combinatorial optimization problems like the Vehicle Routing Problem. % 
\end{enumerate}

\subsection{Validation}

To validate the 
operational utility of the OD data generated by \textsc{MoveOD}, we simulate routing scenarios using VRP algorithms. The VRP problem we solve involves transporting unit demand (commuters) from known origins to known destinations under vehicle-capacity and pickup-time constraints. This formulation directly maps to commuter mobility, with each vehicle representing a shared transport vehicle, such as a shuttle or van.

We construct a validation benchmark by selecting $n = 100$ commuter trips from our calibrated OD data in Hamilton County, Tennessee to obtain $\tilde{\mathcal{A}}$. This sample size provides a practical balance between computational tractability and statistical reliability: it is large enough to capture meaningful trip variability, yet small enough to keep the computational cost of repeated routing experiments manageable. The sampled trips are drawn exclusively from early-morning departures (between 12:00 am and 4:59 am), which serves as a representative subset of the dataset for benchmarking. Each sampled trip $(b_o, b_d, m_o, m_d)$ defines a pickup--delivery pair with unit demand, as described in Section~\ref{sec:instance}.

For every OD instance, we feed the same distance matrix, vehicle capacity, and depot location to each algorithm, cap CPU time at 5\,min, and record the best feasible solution cost.

Since some algorithms (notably Attention and POMO) require offline training, we partition commuter data into disjoint training and testing sets. The split is based on geographic origin zones. To create the test set, we randomly designate a contiguous 20\% of the total CUs as the test region. From these origins, 100 trips are sampled. All other CUs constitute the training region. This provides a realistic generalization test, because learning-based methods must route trips in areas not seen during training.

Each algorithm receives the same inputs: the coordinates of all buildings, cost matrix (shortest path travel times), vehicle capacity of $5$ passengers, and a fleet of $30$ identical vehicles. Algorithms are evaluated on their performance over the \emph{test} set. All results are summarized in Table~\ref{tab:cpdp_results}.

Our experiments show that \MoveOD-generated commuter flows yield tractable, solvable CPDP instances. Solutions are consistent across algorithms, with heuristics such as LKH-3 and Clarke--Wright producing near-optimal routes and learning-based methods generalizing reasonably well. These results support the practical usability of \textsc{MoveOD}'s synthetic demand in routing-centric transportation studies.

% \bibliographystyleapp{ACM-Reference-Format}
% \bibliographyapp{main}

\end{document}